\documentclass[preprint,authoryear,12pt]{elsarticle}

\usepackage{epsfig}

\usepackage{amssymb}
\usepackage{rotating}
\usepackage{lscape}
\usepackage[ps2pdf,%
a4paper=true,%
breaklinks=true,%
colorlinks=true,%
pdfauthor={First Author et al.},%
pdftitle={Template for manuscripts in Advances in Space Research}%
]{hyperref}

\journal{Advances in Space Research}

\begin{document}

\begin{frontmatter}



\title{Delayed Outburst of H 1743--322 in 2003 and relation with its other outbursts}


\author[label1,label2]{S. K. Chakrabarti\corref{cor}}
\cortext[cor]{Corresponding author}
\ead{sandip@csp.res.in}

\author[label1]{D. Debnath}
\ead{dipak@csp.res.in}

\author[label1]{S. Nagarkoti}
\ead{srnagarkoti@csp.res.in}

\address[label1]{Indian Centre for Space Physics, Chalantika 43, Garia Station Rd., Kolkata, 700084, India}
\address[label2]{S. N. Bose National Centre for Basic Sciences, JD--Block, Salt Lake, Kolkata, 700106, India}

\begin{abstract}

The Galactic transient black hole candidate H 1743--322 exhibited a long duration outburst in 2003 after more than 
two and a half decades of inactivity. The 2003 event was extensively studied in multi-wavelength bands by 
many groups. The striking feature is that the total energy released 
is extremely high as compared to that in tens of outbursts which followed.
In this paper, we look at this event and study both the spectral and temporal properties of the source using 
two component advective flow (TCAF) paradigm.
We extract accretion flow parameters for each observation from spectral properties of the decay phase and 
determine the mass of the black hole. We computed
the energy released during all the known outbursts since 2003 and showed that 
on an average, the energy release in an outburst is proportional to the duration of the
quiescent state just prior to it, with the exception of the 2004 outburst.
A constant rate of supply of matter from the companion cannot explain the energy release in 2004
outburst. However, if the energy release of 2003 is incomplete and the leftover is released in 2004, 
then the companion's rate of matter supply can be constant since 1977 till date.
 We believe that erratic behaviour of viscosity at the accumulation radius $X_p$ of matter as well as location the $X_p$ itself, 
rather than the random variation of mass transfer rate from the companion, could be responsible for non-uniformity 
in outburst pattern. We discuss several factors on which the waiting time and duration of the next outburst
could depend.
\end{abstract}

\begin{keyword}
X-Rays:binaries; stars: individual: (H 1743--322); stars:black holes; accretion, accretion discs; radiation:dynamics
\end{keyword}

\end{frontmatter}

\parindent=0.5 cm

\section{Introduction}

The Galactic transient low mass black hole candidate (BHC) H~1743--322 was discovered using HEAO-1 and Ariel-V satellites
 (Doxsey et al. 1977; Kaluzienski \& Holt 1977) around 40 years ago. 
The binary system is at $R.A.=17^{h}46^{m}15.61^{s}$ and $Dec=-32^\circ 14^\prime 0.6^{\prime \prime}$ (Doxsey et al. 1977).
 It is located $8.5 \pm 0.8~ kpc$ away at inclination of $75^\circ \pm 3^\circ$ and its spin is $-0.3 < a < 0.7$ (Steiner et al. 2012).
Multiple studies of various outbursts of this source estimated the mass of H~1743--322 
but dynamically measured value is not available. Petri (2008) estimated
the mass to be $\sim 9-13 M_{\odot}$ based on certain high frequency QPO model.
 Shaposhnikov \& Titarchuk (2009) with the use of another reference mass
estimated its mass at $13.3 \pm 3.2 M_{\odot}$ using their
 spectral index - QPO frequency correlation method. Recently, Molla et al. (2017)  estimated
 its mass to be $11.21^{+1.65}_{-1.96} M_{\odot}$ based on timing and spectral properties of 2010 and 2011 outbursts.
 Bhattacharjee et al. (2017) used the Two component Accretion Flow (TCAF) model to fit data of 2004 outburst and estimated the mass to be in the same range.

 During its 1977-78 outburst, the source was observed several times with the HEAO-1 satellite in hard 
 X-rays (Cooke et al. 1984). After that it was dormant in X-rays for more than two decades. On 2003 March 21, the source 
was rediscovered by the INTEGRAL satellite (Revnivstev et al. 2003). The duration of this outburst was long compared to its other outbursts 
and it was studied by various groups in different wave-bands to explore multi-wavelength properties of the source. 
Though this $2003$ outburst was studied extensively by Parmar et al. (2003); Homan et al. (2005); Remillard et al. (2006a); 
McClintock et al. (2009) in X-rays, McClintock et al. (2009) in optical and near-infrared and Steeghs et al. (2003); Rupen et al. (2003) 
in radio bands etc., it has not been studied with physical accretion flow models such as, TCAF solution to obtain physical flow parameters. 
Furthermore, no attempt has been made to compare all these outbursts in terms of the total energy release, peak counts, 
duration of soft states and duration of outbursts and most importantly to have a general understanding of the behaviour of this source across
the outbursts. 
Earlier, Yu \& Yan (2009) and Yan \& Yu (2015) made comparative studies for several outbursts observed by RXTE. We use a similar method 
to compute the total energy release in this paper from the RXTE/ASM data or MAXI/GSC data and go beyond to understand the accretion flow 
properties and the nature of mass transfer rate from the companion.

 The evolution of the spectral and the temporal properties during 2003 outburst is generally found to be similar to other classical
 outbursting sources, such as GRO~J1655--40, XTE J1550--564, GX 339--4, etc.
 The presence of low-frequency as well as high frequency quasi-periodic 
 oscillations (QPOs) (Homan et al. 2005; Remillard et al. 2006b), other than observation of strong spectral
 variability during this particular outburst of H~1743--322, makes it an ideal case to study in X-rays
 (Capitanio et al. 2005; Homan et al. 2005; Remillard et al. 2006a; Kalemci et al. 2006; Prat et al. 2009;
 McClintock et al. 2009; Stiele et al. 2013).
 Large-scale relativistic X-ray and radio jets were also observed during the 2003 outburst (Rupen et al. 2004; Corbel et al. 2005),
 largest one appeared to have started about two weeks before the peak X-ray flux is reached.

Since the 2003 outburst, several recurring X-ray outbursts of roughly two months duration at every six months to two years gap 
have been observed.
The integrated energy release in the 2003 outburst is several order of magnitude larger as compared to the 
other outbursts. The overall light curve of this event has many high and low counts giving it 
an unsettling look. Using TCAF model, the spectra of
several outbursts of this source which occurred in 2004, 2010 and 2011 (Bhattacharjee et al. 2017; Debnath et al. 2013;
Mondal et al. 2014; Molla et al. 2017) were fitted.
The successful fits allow us to find the variation of the accretion rates of the Keplerian, the sub-Keplerian components of the flow, 
and the size of the Compton cloud on a daily basis. Because the normalization of TCAF (a factor between the computed spectrum and the observed spectrum by a specific instrument) is a contant, even from each independent sample of spectral data, we could estimate the 
mass of the black hole and not surprisingly they agree within acceptable error-bar. We therefore were motivated to fit the spectra
of the 2003 outburst using TCAF which relies on steady state equations of the
viscous (disk) and inviscid (halo) transonic flow. We also want
to generate insight into this system as a whole, as it moved from one outburst to another.

In the next Section, we study the spectral and temporal properties of the X-ray data obtained during the 2003 outburst
using TCAF model. In \S3, we present the flow parameters and the black hole mass obtained from this outburst.
Because of the unsettling nature, conventional phenomenological model diskbb plus power-law gives very fluctuating disc and Compton cloud parameters
during most of the days of this outburst (McClintock et al. 2009). It is possible that the flow accretion rate itself varies with 
radial distance either because some magnetic fields are taking away matter from various radii or because the flow supply from the
accumulation radius itself is erratic and there was always a radial variation of the accretion rate of the Keplerian component. In 
either of these two cases, no existing model is capable of fitting the spectra satisfactorily.
With the physical TCAF solution, which requires steady state flows, fits of 
those days become unsatisfactory, i.e., model fitted reduced $\chi^2$ values were found to be $> 2$.
Hence, we had to choose the data near the end of the outburst where the disc
became somewhat steady. In \S 4, we study all sky monitor data from RXTE and MAXI/GSC of all the outbursts since 2003 and 
compare their behaviour. We study the variations of total energy released per outburst and the amount of matter accumulated in the preceding quiescent state. 
In \S 5, we briefly present a scenario of how the disc may be evolving with time and draw our conclusions.
 
\section{Observation and Data Analysis}\label{sec:observation_data_analysis}

The archival data from RXTE observations using All Sky Monitor (ASM) and Proportional Counter Array (PCA)
unit 2 (PCU2) are used for temporal and spectral study. We also use MAXI/GSC one day average data to
show source behaviour after RXTE era. We use Crab conversion factor of $75$~Counts/sec for RXTE ASM data 
in $1.5-12$~keV energy band and $2.82$~$photons~sec^{-1}~cm^{-2}$ for MAXI GSC data in $2-10$~keV energy range, 
to make count rates in generalized Crab unit as is usually done (see http://xte.mit.edu).

For the timing analysis, we use PCA Science Binned/Event mode data with maximum timing resolution of $125 \mu s$ to
 generate light curves for PCU2 data in 2-15 keV (0-35 channels).
 The routine ``powspec" was used to compute rms fractional variability on 2-15 keV light curves
 of $\sim 0.01$s time bin. We obtain the power density spectra (PDS) after normalization and dead-time
 corrections. The QPO profiles are assumed to be Lorentzian functions and are decided on the basis of
 Q-factor, $Q=\nu/FWHM > 2$. We average QPO properties over each observation and find QPOs for all observations
 available throughout the outburst.
 
We use the spectra obtained from PCU2, the best-calibrated detector
 module in PCA (Jahoda et a. 2006) to carry out the spectral analysis of 2003 outburst data. These spectra are $16s$ binned
 ``Standard2" data. The spectra were corrected for background and a systematic error of 1\% (McClintock et al. 2009)
 was used to account for statistical errors in counts. For the analysis, HeaSOFT package version 6.19 was used with
 XSPEC 12.9.0 (Arnaud 1996). We use TCAF model (local model with {\it fits} file TCAF\_v0.3.fits) following methods explained in
 greater details by
 Debnath et al. (2014, 2015a,b); Mondal et al. (2014, 2016); Molla et al.
(2016, 2017); Jana et al. (2016); Chatterjee et al. (2016); Bhattacharjee et al. (2017)
 to obtain fits of the spectra along with tbabs (Wilms et al. 2000) 
 to account for interstellar extinction. We take the fixed value of hydrogen column density at
 $N_{H}=1.6\times 10^{22}~atoms~cm^{-2}$ (Capitanio et al. 2009). In most of the spectra, we add a Gaussian
 $Fe$ $K\alpha$ emission line to get a better model fit. The central line energy and line width of
 Gaussian were kept in the range of $6.2-7.0$~keV and $0.1-0.8$~keV respectively. 

\section{Results of the 2003 Outburst}\label{sec:results}

\subsection{The Outburst profile}\label{sec:profile}

After its discovery in 1977, the source was completely
 dormant until its high flaring activity in 2003. Although
 a detection ($\sim5~mCrab$) was reported by EXOSAT in 1984 (Parmar et al. 2003; Reynolds et al. 1999), it
was just a passing detection at a quiescence level during slewing the telescope and was
 not an outburst. Another weak X-ray activity was reported by TTM in 1996 (Emelyanov et al. 2000) but RXTE, though 
 active in this era, never mentioned about it. Thus we ignore both of these `events'.
 This outburst started on $21^{st}$ March 2003 (Revnivstev et al. 2003)
 and ended in last week of October 2003 (Tomsick\& Kalemci 2003). Since then, RXTE along with other satellites, 
obtained data from other outbursts of H~1743--322 in $2004,
~2005,~2007,~2008,~2009,~2009/10,~2010 \mathrm{and}~2011$.
In the years after RXTE was decommissioned, H~1743--322 continues to show outbursting
 behavior with a similar duration of $\sim 2$~months with a gap of $\sim 6$~months to $\sim$ year. 
Fig. 1 shows the RXTE/ASM 1.5 - 12 keV light curve (online violet) and the MAXI/GSC 2-10 keV light 
curve (online green) of H 1743--322 during RXTE (starting from Aug. 2002; MJD=52500) and MAXI era (Aug. 2009 to Jan. 2017), respectively.
Usually, maximum RXTE/ASM $1.5-12$~keV count rates during the outbursts of this BHC in RXTE era is $\sim 0.2-0.3$~Crab.
The 2008 outburst of H~1743--322 was dubbed as ``failed", where the count rates did not go above 0.133 Crab ($10$ counts per second) (Capitanio et al. 2009; Zhou et al. 2013)
and softer spectral states were not observed. However, Fig. 1 shows that during 2003 outburst, the maximum ASM
$1.5-12$~keV count rates reached 1.33 Crab ($\sim 100$ counts per second). It is the brightest outburst that H~1743--322 has shown during RXTE era.
Only a few other BH sources have ever reached more than this high value of count rates. The 2003 outburst lasted for
the longest duration of $\sim 8~months$, whereas other H~1743--322 outbursts lasted for $\sim 2~months$. 



\begin{figure}
\centerline{
\includegraphics[width=0.75\columnwidth]{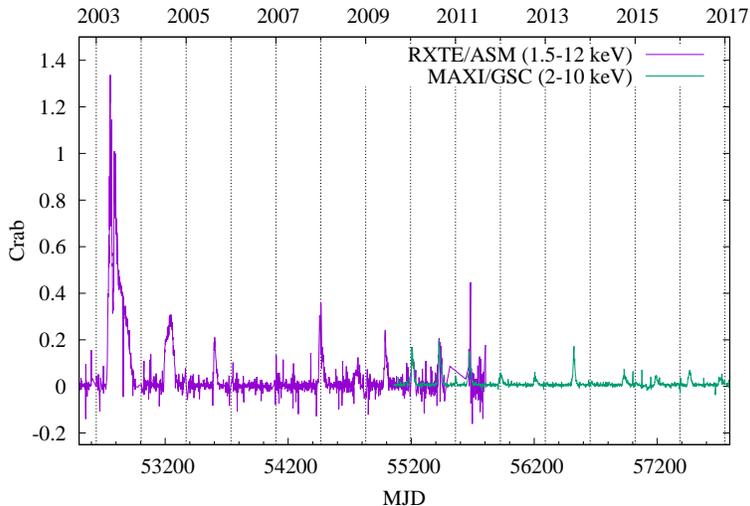}}
\caption{
Daily averaged count rate of (online violet) RXTE/ASM in $1.5-12$~keV (in Crab) and (online green) MAXI/GSC  
in $2-10$~keV (in Crab). 
 \label{fig:asm}}
\end{figure}

It is clear that after the 2003 outburst, H~1743--322 underwent a fundamental change in its accretion disc configuration from which it has not reverted
back till date. 
It did not show any major long outburst for more than two decades before 2003 outburst. However, afterwards, it has shown 
short duration ($\sim 2$~months) outbursts every year. A stable quiescent configuration which existed between $1977/78$ and $2003$ 
outbursts seems to have been disturbed. During the outburst of 2003, the disc could be unsettled as 
no satisfactory fits with conventional models could be made. The source went to the Hard State at the end of 
2003 outburst. This Hard State was found to differ from the pre-outburst Hard State because a disc black body 
component continued to exist even after the Soft State was over (Capitanio et al. 2005;
Lutovinov et al. 2004). This is unusual and warranted investigation as to whether the disc was really evacuated in 2003. 
Following the transition of H~1743--322 to the 
Hard State (Tomsick\& Kalemci 2003) on  MJD 52933, significant radio emissions were also found (Corbel et al. 2005).
We discuss about the possible scenario later below.

In Fig. 2, we show how we fitted each outburst light curve (black lines with points) 
with Fast Rise and Exponential Decay (FRED) profile (online blue). The quiescent states generally have a 
\textbf{count rate equivalent to} about a few mCrab. A limit of $12$mCrab was taken for all the outbursts so that energy release 
above may be integrated to measure the Integrated Counts released per outburst.

The Fast Rise Exponential Decay (FRED) profile was developed by Kocevski et al. (2003) and has since been used 
by many to interpolate the data gaps. The behavior of the light curve is described by the relation,
 $F(t)=F_m \bigg(\frac{t}{t_m}\bigg)^r \bigg[\frac{d}{d+r}+\frac{r}{d+r}\bigg(\frac{t}{t_m}\bigg)^{r+1}\bigg]^{\frac{-(r+d)}{r+1}} $,
 where, $F_m$ is the maximum flux at time $t_m$. The rising and decaying indices are represented by $r$ and $d$, respectively.
 We used this profile to fit ASM/GSC data of all outbursts of H 1743--322. In Fig. 2, we note that 
there are seven peaks in this data, we used a combination of seven FRED profiles to fit this 
outburst behaviour. Once proper fits were obtained, we used quiescent count value ($12 ~mCrab$) to mark
the start and end of the outbursts. In Fig. 2, both start and end are shown with short vertical 
red lines drawn using the above criterion.

\begin{figure}
\centering
\includegraphics[width=0.75\columnwidth]{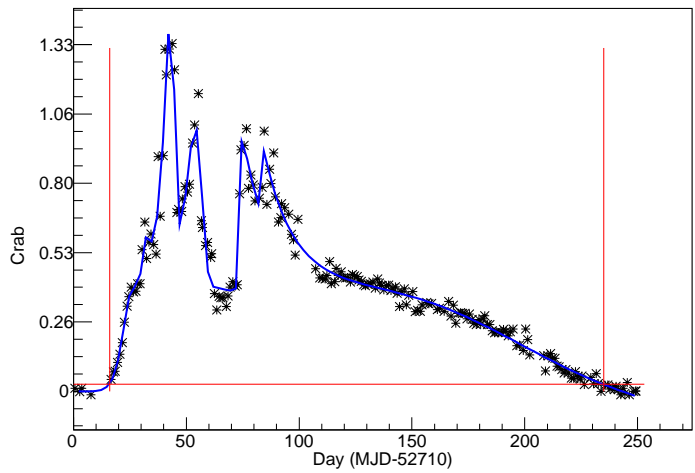}
\caption{FRED profile fitting on 2003 outburst data (blue). For all outbursts, the quiescent
  value of counts was taken to be equivalent to 12mCrab. A horizontal line (red) with this value of
  counts was drawn. The points where the FRED-fitted curve touches this line are the
  start (in the rising phase) and end (in the declining phase) of the outburst. Both start and end
  are shown with short vertical red lines. This method is followed for all outbursts of our study .}
\label{fig:2003_fred}
\end{figure}

\subsection{Spectral Analysis of declining phase of 2003 outburst}\label{subsec:spectral}

We studied the declining HIMS and HS data with the current 
version (v0.3) of the TCAF model fits file in XSPEC. A reasonable good spectral fit (reduced $\chi^{2}<1.2$) 
with the TCAF solution during the declining phase (MJD $52936-50$) was obtained. Fits in other regions 
of the 2003 outburst with the TCAF, were found to be unsatisfactory. There are two possibilities for this:
(a) The accretion rate of Keplerian matter rushing in due to enhanced viscosity at the accumulation radius may develop radial 
dependence at time scales below viscous time because of which no model, which typically assumes steady rates, 
can fit the spectra. (b) Matter with high angular momentum may start shading gas all the way in the form of outflows. As a result a radial dependence 
in accretion rate may ensue which causes failure to fit the spectrum. 
Such issues do not arise for a phenomenological model such as a diskbb plus power-law model which does not question
the origin of the Compton cloud or the seed photons. The rising phase of the outburst is highly jet dominated with large fluctuations 
in disc blackbody and power-law component fluxes. 
The whole profile exhibits multiple-outbursting activities within a single 
`2003' outburst. We interpret this to be due to wild fluctuation of viscosity at the accumulation radius in the outer disc.
 We discuss this later.
 
Five model input parameters, (i) black hole mass ($M_{BH}/M_{\odot}$), (ii) Keplerian disc accretion rate ($\dot{m}_{d}/\dot{M}_{Edd}$), 
(iii) sub-Keplerian accretion rate ($\dot{m}_{h}/\dot{M}_{Edd}$), (iv) the shock location, ($X_s/(r_g=2GM_{BH}/c^2$)) 
and (v) compression ratio at shock ($R$=ratio of post-shock and pre-shock densities) other than normalization are used in TCAF-model. 
The spectral parameters obtained from our analysis are shown in Table 1. 
From the spectral fits, the mass of H~1743--322 is found to be in between $9.4~M_{\odot}$ to $12.5~M_{\odot}$. The average of the observed 
mass of the source is $\sim 11.4 \pm 0.33 M_{\odot}$. TCAF enables us to estimate the mass just by restricting the normalization 
to a constant or nearly constant value (its actual number derived through the fits is not important determining the mass itself). 
The normalization ($Norm$) is found in a narrow range ($0.61-0.67$) 
with an average value of $\sim 0.62 \pm 0.17$. The shock-location ($x_s$) moves outward from $\sim 21 r_g$ to $\sim 372 r_g$. 
The compression ratio (R) increases from $\sim 1.1$ to $\sim 2.3$. To obtain best fit, we used Gaussian Fe-line at $\sim 6.5$~keV 
for most of the observations. The mass obtained from this analysis is in the same ball park as obtained in previous estimates 
mentioned in \S 1. However unlike some other works, our mass measurement is 
stand-alone. 
The evolution of all parameters during the declining phase are shown in Fig. 3. 

\begin{landscape}
\begin{table*}
\addtolength{\tabcolsep}{-3.50pt}
\small
\caption {\leftline{Spectral properties}}
\vskip 0.2cm
\begin{tabular}{cccccccccccc}
\hline
S.N. &Obs-Id   &  MJD      & $\dot{m}_d$      &$\dot{m}_h$      &$M_{BH}$   &$X_s$  &R&Norm.& Gaussian&$\chi ^2_{red}$/d.o.f. \\
     &       & (52900+) & ($\dot{M}_{Edd}$)&($\dot{M}_{Edd}$)&($M_\odot$)&($r_g$)& &     & (keV)   & \\
 & (1)&  (2)  & (3)  & (4)& (5) & (6) & (7) & (8) & (9) & (10)\\
\hline
1  & X-24-00 & 36.047 &  2.49$_{ -0.05 }^{+ 0.05 }$&0.485  $_{ -0.011 }^{+ 0.004 }$& 9.37    $_{ -0.42 }^{+ 0.11 }$&20.4 $_{ -0.3 }^{+ 0.3 }$&   1.08    $_{ -0.02 }^{+ 0.01 }$& 0.62    $_{ -0.02 }^{+ 0.03 }$& 6.95     & 1.19/40 \\
2  & X-25-00 & 37.047 &  2.79$_{ -0.06 }^{+ 0.05 }$&0.477  $_{ -0.019 }^{+ 0.019 }$& 10.23    $_{ -0.66 }^{+ 0.55 }$&26.6 $_{ -1.6 }^{+ 1.2 }$&   1.11    $_{ -0.02 }^{+ 0.02 }$& 0.60    $_{ -0.01 }^{+ 0.01 }$& 6.92     & 1.14/40 \\
3  & X-26-00 & 38.016 &  2.58$_{ -0.03 }^{+ 0.04 }$&0.456  $_{ -0.014 }^{+ 0.010 }$& 13.97    $_{ -0.50 }^{+ 0.86 }$&32.4 $_{ -1.8 }^{+ 1.4 }$&   1.08    $_{ -0.02 }^{+ 0.02 }$& 0.61    $_{ -0.01 }^{+ 0.03 }$& 7.00      & 1.16/40 \\
4  & X-27-00 & 39.133 &  2.37$_{ -0.05 }^{+ 0.04 }$&0.423  $_{ -0.006 }^{+ 0.006 }$& 11.92    $_{ -0.46 }^{+ 0.48 }$&31.1 $_{ -0.8 }^{+ 1.9 }$&   1.08    $_{ -0.01 }^{+ 0.01 }$& 0.59    $_{ -0.01 }^{+ 0.01 }$& 6.83     & 1.00/40 \\
5  & X-28-00 & 42.156 &  1.81$_{ -0.05 }^{+ 0.06 }$&0.336  $_{ -0.005 }^{+ 0.004 }$& 12.18    $_{ -0.86 }^{+ 0.90 }$&42.9 $_{ -1.1 }^{+ 1.2 }$&   1.09    $_{ -0.02 }^{+ 0.04 }$& 0.62    $_{ -0.05 }^{+ 0.05 }$& 6.72     & 0.57/40 \\
6  & X-29-00 & 42.703 &  1.51$_{ -0.06 }^{+ 0.03 }$&0.292  $_{ -0.025 }^{+ 0.045 }$& 12.15    $_{ -0.18 }^{+ 0.18 }$&47.3 $_{ -1.9 }^{+ 2.1 }$&   1.09    $_{ -0.02 }^{+ 0.04 }$& 0.61    $_{ -0.01 }^{+ 0.01 }$& ---       & 0.84/43 \\
7  & Y-01-00 & 44.137 &  1.12$_{ -0.08 }^{+ 0.04 }$&0.266  $_{ -0.016 }^{+ 0.013 }$& 11.26    $_{ -0.70 }^{+ 0.76 }$&86.0 $_{ -5.1 }^{+ 7.5 }$&   1.08    $_{ -0.02 }^{+ 0.01 }$& 0.62    $_{ -0.04 }^{+ 0.04 }$& 6.55     & 0.70/40 \\
8  & X-30-00 & 45.180 &  0.88$_{ -0.03 }^{+ 0.03 }$&0.240  $_{ -0.013 }^{+ 0.013 }$& 12.54    $_{ -0.12 }^{+ 0.54 }$&136.5 $_{ -8.6 }^{+ 8.6 }$&   1.11   $_{ -0.02 }^{+ 0.01 }$&  0.64    $_{ -0.06 }^{+ 0.06 }$& 6.66     & 0.59/40 \\
9  & Y-01-01 & 46.105 &  0.74$_{ -0.03 }^{+ 0.06 }$&0.231  $_{ -0.040 }^{+ 0.014 }$& 10.99    $_{ -0.44 }^{+ 0.47 }$&179.2 $_{ -2.3 }^{+ 2.3 }$&   1.33   $_{ -0.03 }^{+ 0.04 }$&  0.62    $_{ -0.03 }^{+ 0.03 }$& 6.74     & 1.11/40 \\
10 & X-31-00 & 47.043 &  0.70$_{ -0.01 }^{+ 0.02 }$&0.199  $_{ -0.018 }^{+ 0.018 }$& 10.49    $_{ -0.12 }^{+ 0.13 }$&222.3 $_{ -6.2 }^{+ 6.4 }$&   1.60   $_{ -0.08 }^{+ 0.08 }$&  0.67    $_{ -0.06 }^{+ 0.06 }$& 6.70     & 0.89/40 \\
11 & X-32-00 & 48.211 &  0.65$_{ -0.02 }^{+ 0.02 }$&0.173  $_{ -0.008 }^{+ 0.009 }$& 11.33    $_{ -0.65 }^{+ 0.71 }$& 245.5 $_{ -5.2}^{+14.4 }   $& 1.67  $_{ -0.08 }^{+ 0.09 }$&   0.63    $_{ -0.04 }^{+ 0.04 }$& 6.65     & 0.91/40 \\
12 & Y-01-02 & 49.195 &  0.59$_{ -0.02 }^{+ 0.03 }$&0.149  $_{ -0.008 }^{+ 0.003 }$& 11.17    $_{ -0.68 }^{+ 0.78 }$&299.1 $_{ -3.9 }^{+3.9 }  $&   1.97   $_{ -0.09 }^{+ 0.10 }$&  0.66    $_{ -0.04 }^{+ 0.05 }$& 6.62     & 0.88/40 \\
13 & X-34-00 & 50.457 &  0.53$_{ -0.02 }^{+ 0.02 }$&0.102  $_{ -0.007 }^{+ 0.002 }$& 12.41    $_{ -0.17 }^{+ 0.17 }$&372.3 $_{ -9.9 }^{+16.0 } $&  2.31 $_{ -0.08 }^{+ 0.08 }$&    0.64    $_{ -0.04 }^{+ 0.03 }$& 6.69     & 0.95/40 \\
\hline
\end{tabular}
\vskip 0.5cm
{\leftline{X=80137-01; Y=80137-02}}
\label{tab:spec2}
\end{table*}
\end{landscape}
 
The state transition from the Hard-Intermediate State to Hard State which occurred on $MJD~52942.1$ is indicated by a vertical
line in Fig. 3. This is done on the basis of changes in shock location and compression ratio. 
Since TCAF analysis covers only the declining HIMS and HS regions of the outburst, we rely on spectral and temporal 
analysis of McClintock et al. (2009) using combined disc blackbody, power-law and broken power-law models. Based 
on the degree of importance of thermal (disc black body) and non-thermal (power-law or broken power-law) 
components and nature of QPOs (if present), as suggested by Debnath et al. (2013), we can classify the entire 
2003 outburst into four spectral states, Hard (HS), Hard-Intermediate (HIMS), Soft-Intermediate (SIMS) and 
Soft (SS). The nature of the variations of hardness ratios also 
confirms our classification. As of other transient BHCs (GRO~J1655--40, GX~339--4, MAXI~J1659--152, MAXI~J1543--564) reported previously 
(see, Chakrabarti et al. 2008; Nandi et al. 2012; Debnath et al. 2008, 2015a,b; Chatterjee et al. 2016), 
monotonic evolutions of 
QPO frequencies are observed in HS and HIMS from both the rising and declining phases of the outburst and sporadic QPOs 
are seen during SIMS from both rising and declining phases of the outburst and no low frequency QPOs are observed in SS. 
It seems that spectral transition occurs from HS (Rising phase) to HIMS (Rising phase) on 2003 March 31 (MJD=52729.8), on 2003 April 6 
(MJD=52735.7) from HIMS (Rising phase) to SIMS (Rising phase), on 2003 June 30 (MJD=52830.4) from SIMS (Rising phase) to SS, on 2003 October 18 
(MJD=52930.9) from SS to SIMS (Declining phase), on 2003 October 24 (MJD=52936.0) from SIMS (Declining phase) to HIMS (Declining phase) and finally on 
2003 October 30 (MJD=52942.7) from HIMS (Declining phase) to HS (Declining phase). This last transition is marked with a vertical line in Fig. 3.

\begin{figure}
\centerline{
\includegraphics[width=0.75\columnwidth]{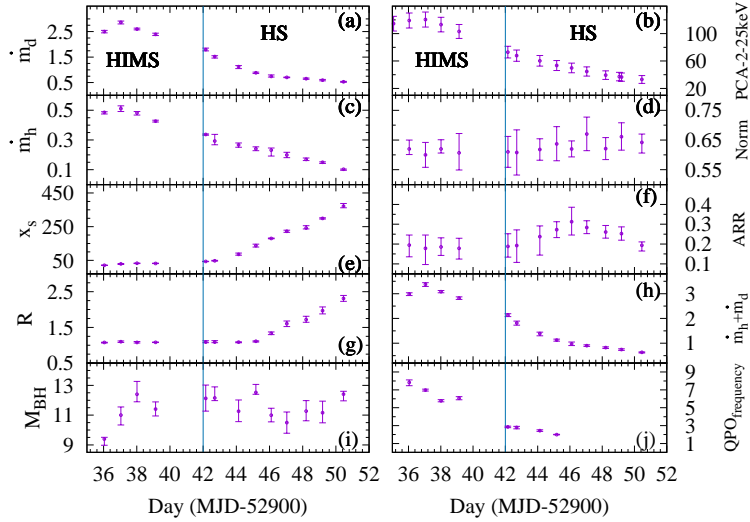}}
\caption{Showing results from TCAF fits.
 Variation of (a) $\dot{m}_{d}$ (b) PCA count rates (2-25 keV) (c) $\dot{m}_h$ (d)Normalization (e) $X_s$ 
 (f) Accretion Rate Ratio ($ARR=\dot{m}_h / \dot{m}_d$) (g) $R$ (h) $\dot{m}_d$+ $\dot{m}_h$ (i) $M_{BH}$ and (j) $QPO_{freq}$ 
  with MJD. \label{fig:tcaf}}
\end{figure}

\subsection{Study of QPO evolution with POS Model}\label{subsec:pos}

In TCAF paradigm, the shock oscillations and consequent oscillations of Comptonized X-rays result in 
Low Frequency QPOs (Chakrabarti \& Manickam 2000). Type-C QPOs are the results of near resonance between 
cooling time and infall (compressional heating) time of the post-shock region (Molteni et al. 1996; Chakrabarti et al. 2015)
or non-fulfillment of Rankine-Hugoniot conditions (Ryu et al. 1997). Weak resonances result in the formation of type-B QPOs. 
When shocks are not produced but the centrifugal barrier is formed, and the oscillations are tentative, different regions 
oscillating in different phases, form type-A QPOs. The declining phase of the 2003 outburst of H~1743--322 shows
signatures of type-C QPOs and transitions from HIMS to HS as seen in other outbursts (Debnath et al. 2013,
2015a; Molla et al. 2016). We study the evolution of QPO frequency during this declining phase 
using the so-called propagating oscillatory shock (POS) model (Chakrabarti et al. 2005, 2008; Debnath et al. 2010, 2013; Nandi et al. 2012).
In this case, the shock is propagating and oscillating at the same time, giving rise to variation of QPO
frequencies since both the time scales are also changing.
 The equation for QPO frequency in this model is $\nu_{QPO}=\beta/[X_s(X_s-1)^{1/2}]$ where `$\beta=1/R_0\pm t_d^2 \alpha$' is
 the shock strength, `$R_0$' is the compression ratio `$R$' on the first day of QPO evolution, $t_d$ is
 the time in days and $\alpha$ is a constant number deciding how `$R$' strengthens or weakens with
 QPO evolution period. The instantaneous shock location is given as, $X_s(t)=X_{s0}\pm V(t)t_{d}/M_{BH}$ where
 $X_{s0}$ is the shock location on the first day, `-' sign is used for rising phase 
 when shock comes closer to the black hole and `+' sign is used for the declining phase when the shock location increases with time.
The instantaneous velocity of shock is given by $v(t)=v_0\pm ft_d$ where $v_0$ is the velocity on 
the first day and `$f$' is the acceleration/deceleration of the shock.

\begin{figure}
\centerline{
\includegraphics[width=0.75\columnwidth]{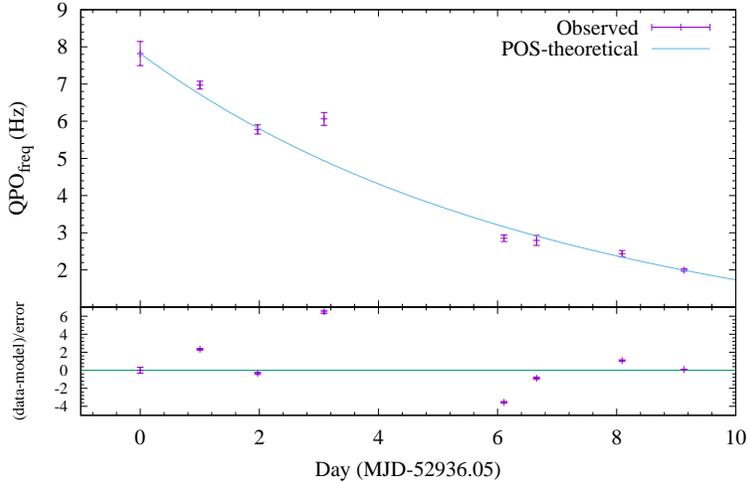}}
\caption{$QPO_{freq}$ evolution during the declining phase, fitted with the POS model.
\label{fig:pos4}}
\end{figure}

We study the QPO evolution in the declining phase of the 2003 outburst from October 24 (start of HIMS) till 
November 2 (the last day when QPO was observed) for the duration of 9 days. A best fit of the evolution was obtained 
using the average mass which was found from TCAF model present above with a slowly accelerating ($f=8~cm~s^{-1}~day^{-1}$) shock wave. 
The shock starts to move outwards on $MJD=52936.05$ from $\sim 73 r_g$ at $\sim 294~cm/sec$ and reaches $\sim 158~r_g$ 
at $367cm/sec$ on the last day when QPO was observed. This is normal and is seen in all other outbursting objects. 
Fig. 4 shows the theoretical values of QPO frequencies plotted 
along with observed values. Table 1 shows the POS-model fitted parameters during the QPO evolution.

\begin{table*}
\addtolength{\tabcolsep}{-3.50pt}
\centering
\small
\caption {\leftline{QPO evolution fitting with POS model}}
\vskip 0.2cm
\begin{tabular*}{\textwidth}{c @{\extracolsep{\fill}} cccccccc}

\hline
S.N.& Obs-Id& MJD   & $QPO_{freq}$& $QPO_{freq}$& $X_s$ & v      &  R \\
    &       &       & Observed    & POS        &      &          &\\ 
    &       & (Days)& (Hz.)      & (Hz.)      & ($r_g$) & ($cm/s$) & \\
 (1)&  (2)  & (3)  & (4)& (5) & (6) & (7) & (8)\\
\hline
1 & X-24-00 & 52936.0 & $7.82{\pm 0.32}$ &  7.82 &    73.7  & 300 &  1.78   \\
2 & X-25-00 & 52937.0 & $6.97{\pm 0.11}$ &  6.72 &    81.3  & 308 &  1.79  \\
3 & X-26-00 & 52938.0 & $5.83{\pm 0.12}$ &  5.82 &    89.1  & 315 &  1.80  \\
4 & X-27-00 & 52939.1 & $5.86{\pm 0.17}$ &  4.94 &    98.5  & 324 &  1.82  \\
5 & X-28-00 & 52942.2 & $2.85{\pm 0.09}$ &  3.16 &   126.6  & 348 &  1.95  \\
6 & X-29-00 & 52942.7 & $2.80{\pm 0.14}$ &  2.92 &   132.0  & 353 &  1.99  \\
7 & Y-01-00 & 52944.1 & $2.44{\pm 0.09}$ &  2.34 &   147.0  & 364 &  2.10  \\
8 & X-30-00 & 52945.2 & $2.00{\pm 0.03}$ &  1.99 &   158.4  & 373 &  2.21  \\

\hline
\end{tabular*}
\vskip 0.5cm
{\leftline{X=80137-01; Y=80137-02}}
\label{tab:pos3}
\end{table*}


\section{Comparison among the Outbursts of H1743--322}

During our analysis, we found that the 2003 and part of 2004 (see also Bhattacharjee et al. 2017) outbursts are
 very unsettling in the sense that either the model fitted parameters fluctuate on a daily basis as in the phenomenological 
diskbb plus power-law model or it is impossible to fit the data with TCAF which satisfactorily fitted 
other outbursts (Mondal et al. 2014; Molla et al. 2017). Thus the disc structure is clearly non-steady and evolving. 
Therefore, we wanted to compare all the outbursts observed till date and and draw some conclusions about the 
general scenario about disk evolution and supply rate of matter from the companion. 

At the outset we must mention that the black hole outbursts are totally different from the dwarf novae outbursts
in that the high energy radiation observed in the former clearly indicates interaction of the Compton cloud
with the disc, and both could be emitting X-rays. It is also clear that for black hole candidates,
the so-called Compton cloud must change its property very fast and this is possible if there is a steady supply of sub-Keplerian and
hot halo matter which makes and constantly replenishes the Compton cloud. TCAF model is based on exact solutions of 
transonic flows whose topologies change fundamentally at a critical viscosity parameter. For sub-critical parameter,
the flow still forms shocks close to the centrifugal barrier as in an inviscid flow (Chakrabarti, 1990). However, for a super-critical 
viscosity, the barrier is removed and a Keplerian disc passing through the inner sonic point outside the horizon is formed instead.
In TCAF, the formal component with low angular momentum and viscosity surrounds the Keplerian disc with high angular momentum
	and viscosity located at the equatorial plane (Chakrabarti, 1995). This TCAF configuration was later used by Chakrabarti (1997 and references therein)
to study the spectral properties. This configuration has been verified  to be stable by 
extensive numerical simulation of Giri \& Chakrabarti (2013). Since the Compton cloud is the post-shock region where the 
Keplerian disc is also truncated, TCAF uses a minimum set of parameters to fit the spectra. 
On the contrary, in a dwarf novae outburst, both components need not be present
simultaneously. In fact the accretion may switch from high viscosity flow to low viscosity flow due to limit cycle behavior
(e.g., Lasota 2001). The main reason is that in the black hole accretion, the outgoing radiation 
is emitted from a few hundred Schwarzschild radius where the infall time is much shorter compared to the
viscous time scale. On the contrary, the size of a white dwarf could be about 3000 Schwarschild radius and the 
physics of novae is decided by what happens very very far at a much cooler disc. The outbursts in both the cases are dictated by 
viscosity. If viscosity is not enough, matter cannot proceed towards the compact object and is piled up at some location 
(pile-up radius or accumulation radius) away 
from the black hole, say at $X_p$, till some instability raises the viscosity to drive the piled up matter towards the compact object 
and causing the outburst. Thus the piling radius is a function of viscosity available and could change from one outburst to another. 
We believe that this is precisely what is happening in the object under study. 

Another point to remember is that during an outburst, the energy that is released is primarily 
from the gravitational energy of accreting matter. Since matter is piled up at $X_p$ and not accreted during the quiescent states,
the emission is weaker in the quiescent state and it catches up during the outburst. 
Our main goal would be to see if the net energy released in a given outburst is proportional to the net amount of 
matter that is supplied by the companion during the period 
in between two outbursts (from peak flux of one till the peak flux of the previous one). 
Given that the spectral features did not change from one outburst to another
(i.e., same fraction of gravitational energy release is observed in the same energy range), the
observed energy release in our specific band may be assumed to be proportional to the net energy release 
for each of the outbursts.

\subsection{ASM and GSC data of H~1743--322}

In order to compare all the outbursts, we obtained the total integrated counts (I.C.) observed during each
 outburst with RXTE/ASM 
and MAXI/GSC modified in `Crab sec' unit with appropriate conversion factors. From Fig. 1 it is clear
that in the light curve from MAXI/GSC, the profiles are well defined. So, the outburst of 2010, which
 is common in both RXTE/ASM and MAXI/GSC, was used as a reference (normalization of I.C.) 
while comparing the events from two instruments. Figure 5a shows the I.C. per day of outburst, normalized with 
respect to 2010 event (Normalized counts per day of outburst) as shown in Table 2. The area of each 
histogram represents the I.C. while the 
width represents the duration. We clearly see that the 2003 outburst is stronger by orders of magnitude
than any other outburst which followed. Generally, the I.C. shows a gradual decrease. 
This can be easily converted into total energy release rate (see also, Yan\& Yu 2015) using the 
prescription (see http://xte.mit.edu) and taking proportional energy after conversion from RXTE to MAXI. 
 
\begin{figure}
\centering
\includegraphics[width=0.75\columnwidth]{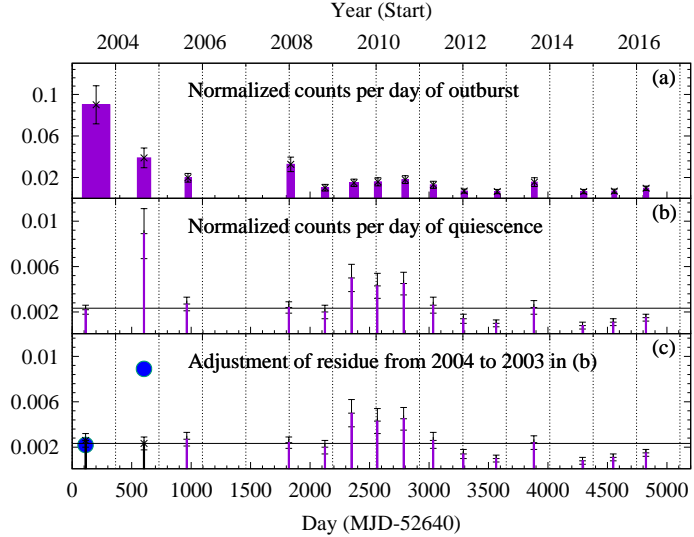}
\caption{(a) The I.C., normalized with I.C. in 2010 outburst received by the same instrument, i.e., by ASM or GSC.
per day of outburst. All data before 2010 are from ASM and the rest are from GSC.
	(b) The I.C. normalized as before with I.C. of 2010 per day of preceding period (peak flux to peak flux). For
 2003 outburst, we considered previous outburst to be in 1977. (c) Same as Panel b, except that the excess I.C. of the
 2004 outburst, over the average I.C. (shown by horizontal solid line) is transferred into the 2003 outburst.
 Data without this transfer (as in the middle panel) are shown with circles.}
\label{fig:figure_fred}
\end{figure}

In Fig. 5b, we show the I.C. per day of the preceding period (measured as 
the peak flux to peak flux of two successive outbursts) of each outburst with day. 
We note that the energy release rate of the 2004 outburst is very high for the short 
accumulation period as compared to the other outbursts. Interestingly, if for 
the sake of argument, we assume that 2004 outburst also releases a part of the energy not released by
the 2003 event, then we can distribute the total of 2003 and 2004 energy emission rate evenly from 1977 till 2004. 
In that case, we find that the average emission due to 2004 outburst comes down to the level where all other 
outbursts generally reside (Fig. 5c). One of the reasons why we felt that 2004 outburst is anomalous is that its 
$\sim 115$ days of duration is much larger than the durations of all outbursts afterward and much of the data 
cannot be fitted by any conventional models, indicating a non-steady flow. Furthermore, as mentioned in the 
introduction, there appeared to be a leftover Keplerian disc even after the 2003 outburst was over. So, may be,
the 2003 outburst was prematurely halted for lack of viscosity and restarted again in 2004 when the 
condition was more favourable.

\begin{landscape}
\begin{table*}
\centering
\small
\caption {\leftline{Peak data and Ratio of integration counts (I.C.) with I.C. of 2010, during various outbursts of H~1743--322}}
\vskip 0.2cm
\begin{tabular}{cccccc}
\hline
Year                                   & Peak Day & Peak Count & Duration    & Quiescent$^\dagger$ & Ratio of I.C.$^{\dagger\dagger}$ \\
                                       &          & rate        &   &  Period             &                         \\
                                       &  (MJD)   &  (Crab)      & (Days)       &  (Days)             & \\
\hline
ASM &  &  &  &  &\\
\hline
      1977                              &  43404.0  &  0.730 $\pm$ 0.09  & ---&  ---  &   ---   \\
      2003$^{\dagger \dagger \dagger}$  &  52752.5  &  1.421 $\pm$ 0.162 &  230.5 &  2635.5  &  20.75 $\pm$  4.22   \\
      2004                              &  53243.7  &  0.295 $\pm$ 0.008 &  112.2 &  491.2  &   4.37 $\pm$  1.07   \\
      2005                              &  53603.2  &  0.189 $\pm$ 0.012 &   49.5 &  359.5  &   0.98 $\pm$  0.21   \\
      2007                              &  54461.4  &  0.260 $\pm$ 0.093 &   62.8 &  858.2  &   2.05 $\pm$  0.44   \\
      2008                              &  54764.7  &  0.093 $\pm$ 0.014 &   60.0 &  303.3  &   0.61 $\pm$  0.19   \\
      2009                              &  54988.2  &  0.224 $\pm$ 0.012 &   74.5 &  223.5  &   1.12 $\pm$  0.26   \\

\hline
GSC &  &  &  & \\
\hline
    2009-10  &  55205.8  &  0.171  $\pm$0.008&    59.3  &  217.6  &   0.94 $\pm$  0.23   \\
      2010   &  55426.6  &  0.187  $\pm$0.019&    55.5  &  220.8  &   1.00 $\pm$  0.21   \\
      2011   &  55675.5  &  0.141  $\pm$0.020&    50.0  &  248.9  &   0.65 $\pm$  0.17   \\
    2011-12  &  55929.0  &  0.055  $\pm$0.004&    51.1  &  253.5  &   0.36 $\pm$  0.10   \\
    2012-13  &  56204.6  &  0.045  $\pm$0.007&    43.3  &  275.6  &   0.28 $\pm$  0.09   \\
      2013   &  56521.7  &  0.173  $\pm$0.015&    48.0  &  317.1  &   0.75 $\pm$  0.20   \\
    2014-15  &  56931.0  &  0.073  $\pm$0.009&    52.3  &  409.3  &   0.34 $\pm$  0.11   \\
      2015   &  57189.4  &  0.045  $\pm$0.002&    40.5  &  258.4  &   0.28 $\pm$  0.08   \\
      2016   &  57463.7  &  0.060  $\pm$0.006&    43.5  &  274.3  &   0.42 $\pm$  0.09   \\
\hline
\end{tabular}
\leftline {$^\dagger$ Quiescent period is the time duration from peak day of previous outburst }
\leftline{to the concerned outburst.}
\leftline{$^{\dagger \dagger}$ Ratio of Integrated Count (I.C.) in each outburst with that of 2010 }
\leftline{ of corresponding instrument (ASM or GSC).}
\leftline{Dashes have been put where data was not available.}
\leftline{Peak values are obtained by fitting Lorentzian profile to the peak of each outbursts.}
\leftline{Data of 1977 are taken from Kaluzienski \& Holt, 1977}
\leftline{$^{\dagger \dagger \dagger}$ Although quiescent period from 1996 to 2003 is 2635.5, we use the period}
\leftline{from 1977 to 2003 in Fig. 5 after ignoring 1996 outburst as it was not detectable by ASM.}
\label{tab:peaks4}
\end{table*}
\end{landscape}

\section{Discussion and Conclusions}\label{sec:discussion}

The outbursting black hole candidate H 1743--322 is intriguing for several reasons. It remained more or less in quiescent state
for about twenty five years till 2003 when it emitted a huge amount of energy for a period of several months. This type of `mega' event 
has not been repeated since then, although outbursts are seen quite regularly. The 2003 outburst has a strange feature 
that it has been impossible to fit the data with a steady disc rate for almost the entire outburst except towards the end when the
disc is settling down. The following outburst of 2004 also showed such un-settling behaviour with a high emission of 
energy and the data could be properly fitted only in harder states (Bhattacharjee et al. 2017). Since then all 
the outbursts are generally `well behaved' and the data could be fitted throughout (Mondal et al. 2014; Molla et al. 2017).

\begin{figure}
\centerline{
\includegraphics[width=0.75\columnwidth]{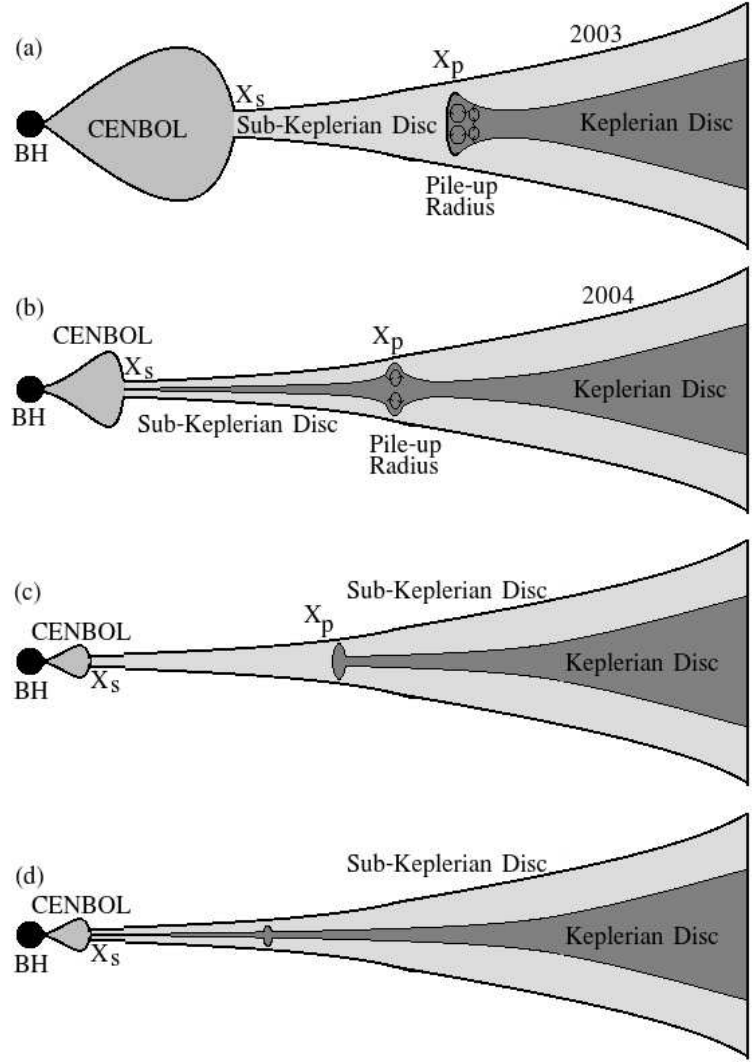}}
\caption{
A possible scenario of the evolution of the disc structure around H 1743--322 in successive
outbursts. (a) The piling radius $X_p=X_{p2003}$ is very high prior to 2003 outburst and thus
considerable matter had to be accumulated before high viscosity released the matter and triggered 
the outburst. (b) Location of piling radius $X_p=X_{p2004}<X_{p2003}$ of 2004 outburst where some matter
of previous outburst remained while the Keplerian disc from 2003 outburst is still fading away. (c) Piling radius at subsequent 
outburst is closer to the black hole and takes shorter time before new triggering occurs (d).}
\label{fig:piling}
\end{figure}

In order to understand the general behaviour of the outbursts of this source, we first recall the limit cycle model 
of dwarf novae outbursts (Cannizzo, 1993) and discuss further.
It is assumed that the accreting gas accumulates in quiescence and then suddenly accretes onto the 
central object during outbursts (e.g., Cannizzo, 1993). It is believed that changing of the phases is 
due to propagation of heating and cooling fronts which traverse the disc and cause phase transitions 
between low (neutral) and high (ionized) states. The column density changes drastically in these two phases. 
It is also believed that the low to high state transition can take place at any distance from the 
compact object (Cannizzo 1998). 
Earlier, we mentioned that in outbursting black hole sources the main triggering is done by enhancement 
of viscosity (Chakrabarti et al. 2005, Debnath et al. 2015a \& Jana et al. 2016 and references therein). 
The process most likely follows this sequence: in low mass X-ray binaries (LMXBs) where the companion supplies matter
through Roche lobe overflow, cannot proceed deep inside due to lack of viscosity. It starts to 
pile up at  $X_p$ dictated by the prevailing viscous processes. 
At this low density quiescence states, before the outburst is triggered,
the advective flow has a shock while the Keplerian disc remains at $X>X_p$. 
The piled up matter heats up the flow on the equatorial plane till convective instability sets in to increase a large viscosity.
Alternatively, magnetic fields advected by the flow from the Companion are anchored at the piled up matter
and could create a large wind/jet and remove angular momentum at the same time. Thus the material
is released and starts to rush in. Indeed in 2003, a strong radio emission was observed much before the
peak X-ray flux was achieved. However, the disc on the equatorial plane remains Keplerian as long as the 
viscosity is supercritical (Chakrabarti, 1990). Otherwise, further formation of the Keplerian disc is halted
and the outburst may not achieve the soft-Intermediate state and soft-state. When viscosity is totally turned off
due to release of most of the stored matter at $X_p$, the outburst starts to decline and eventually reaches the quiescence state.
However, the destruction of a Keplerian disc takes much longer time than its formation, since the two processes
are very much different (Roy \& Chakrabarti, 2017). During the formation of the Keplerian
disc in the rising phase, viscous transport of angular momentum takes place. However, destruction of the Keplerian 
disc takes place by its slow mixing with the advective low angular flow which is much longer.
By this time, the viscosity at $X_p$ may again increase and
cause a further outburst. We suspect that the 2004 outburst was due to such a process. 
The duration soft-intermediate state in the rising phase is decided by the viscous time 
to create the Keplerian disc. Thus the duration of the soft-intermediate states in the rising and declining 
phases along with the presence of a soft state is a good indicator of the size of the Keplerian disc. The radiation 
emitted in these states is also an indicator of the total energy released, since these are the brightest 
states of an outburst. The processes mentioned above is valid only for 
LMXBs. For a high mass X-ray binary (HMXB), outbursts are not possible, since the angular momentum of winds is low and 
does not have to pile-up at a large $X_p$ and store huge matter outside the star 
in the absence of viscosity except at low  $X_p$ where the centrifugal barrier is 
formed even with a low angular momentum. This is the reason why LMXBs exhibit outbursts while HMXBs such as 
Cyg X-1 only show some short-lasting flares when soft states are achieved.

Total amount of accumulation of matter at $X_p$ is guided by two physical processes: the steady transfer 
of matter from the companion through the Roche Lobe superposed by anomalous rate due to the intrinsic 
(magnetic or otherwise) activity of the companion. The total energy released in an outburst is an indicator 
of the total matter accumulated and thus should generally be proportional to the peak-to-peak time gap 
between two outbursts. Any deviation might be the result of variation in supply 
rate of matter due to companion's activity.

After a careful scrutiny of the behaviour found in the data from outbursts of 2003 and 2004, one 
can now construct a scenario of what could have happened 
to the source during these years. A possible sequence is shown in Fig. 6. In 2003 outburst, $X_{p2003}$ must be far out in 
the disc (Fig. 6a) where the flow is neutral and required a huge accumulation of matter in order to raise 
the temperature of the equatorial plane so that the resulting convective instability and the viscosity due 
to turbulence eventually can trigger the outburst after twenty five odd years. Most of this outburst phase 
is unsettling due to rush of matter causing variation of rates with radius. There 
could have been profuse outflows at $X_p$ and outer disk as well to remove angular momentum.
The accretion rate of the disc remained highly radial dependent and thus models which assume a fixed accretion rate cannot fit the 
spectra. From the nature of the variation of spectral and temporal properties during the 2003 outburst, we infer that 
a stable two-component configuration does not form until 2003 October 24 (MJD=52936), i.e., before the start 
of the HIMS (declining). From the behaviour of 2004 outburst, we find that it also does not show any steady 
disc except towards the end. It appears that some matter leaving $X_{p2003}$ during 2003 event could not reach 
the inner disc but remained stuck at a new piling radius $X_{p2004}<X_{p2003}$ possibly due to sudden fall 
of viscosity below critical value while the declining phase of 2003 outburst was still in place. Some more matter was accumulated 
here to trigger the 2004 outburst (Fig. 6b). This also proves that during the so-called quiescent 
state after 2003, the matter supply from the companion continues at a more or less constant rate and a weak 
Keplerian disc continues to emit a multicolor black body (Fig. 6b). Indeed, presence of such a disc around 
H~1743--322 in quiescence has been reported before (Capitanio 2005). Also, 
reduction of viscosity causes the Keplerian disc to dissipate very slowly (Roy \& Chakrabarti, 2017).
Subsequent to 2004, accumulation of matter could be taking place at even closer $X_p$ where the disc is hotter and 
ionized and does not require a large accumulation of matter to trigger the outburst (Fig. 6c). They last for 
$\sim 2$ months with a quiescence period of $\sim 6$~months to $\sim 2$~years. We also observed from 
Fig. 5a, a general trend of weakening of the outbursts as time progresses (Fig. 6d). It is 
possible that after a few such outbursts, it would again remain inactive for two-three decades. The origin 
of this super-cycle of outbursts remains unclear and could be related to the properties of the companion.

There are many comparative studies of various episodes of outburst of an object (e.g., Yan \& Yu 2015 and 
references therein), though no detailed analysis regarding disc flow properties were made. However, we felt that 
combining the data of RXTE/ASM and MAXI/GSC one could have an idea of 
matter transfer rate from the companion. What we established is that if we assume that every flare essentially cleans
up the disc and removes the accumulated matter during the preceding period between the peak flux days, then we cannot explain the 
huge energy release of 2004 outburst unless we assume that the matter supply rate, only in the 2003-2004 outburst time gap, 
was unusually high, which we have no reason to believe.
We propose that 2004 outburst released part of the energy which was due to be released in 2003 itself and perhaps 
because of sudden decline of disc viscosity, the 2003 outburst stopped prematurely. Thus, for example, if we combine
the energy release at 2003 and 2004 and treat them together, the average energy release rate  becomes almost
constant. Other outbursts till date also follow a similar rate, though there were some fluctuations 
presumably due to other effects, such as the mass ejections from the companion, or ablation of matter from the companion 
due to irradiation of X-rays onto the companion, etc. The accumulation radius could be decreasing from one outburst to the next,
as is evidenced from the progressively weaker outburst profile. 
We believe that a similar behaviour should be found in other recurring transient BHCs, such as 
GX~339--4, 4U~1630--472, V404 Cyg etc. where several outbursts have been observed. The determination of the 
detailed dynamics is outside the scope of this paper and will be discussed elsewhere.

In this paper, we presented several new results. We not only analyzed the 2003 declining phase data 
with TCAF model, and obtained the black hole mass independently from 
each fit using the consideration of constant normalization in TCAF,
but also came to a general conclusion regarding the supply rate of the companion which appears to be constant. 
From our analysis of the spectral properties of the declining phase of the outburst with TCAF solution we find 
no unusual behaviour. The mass of the black hole obtained from the present analysis is quite 
consistent with the value proposed by Petri (2008) and others (Shaposhnikov \& Titarchuk
2009; Molla et al. 2017; Bhattacharjee et al. 2017). One important property of fitting with TCAF
is that the normalization $N$ in a fit, is supposed to be a constant for a steady 
two component disc. This normalization is the ratio of the derived spectrum and the observed spectrum by a specific instrument
and is a function of intrinsic property of the system, such as the mass of the black hole, distance of the object and the
inclination angle of the disc. We find that in 2003, $N$ is $\sim 0.6$. However, for 2004 outburst, $N\sim 13$ (Bhattacharjee et al. 2017) 
which is close to what Molla et al. (2016) found for 2010 and 2011 outbursts. This gives us another proof that the  
basic disc properties may have changed from an unsettling in 2003 to a settling nature by the end of 2004 outburst. However, since the mass appears to be in the same ball park in all the outbursts, it proves that the derived mass itself
it not dependent on the exact value of the normalization.

\section*{Acknowledgments}

This research has made use of archival data of RXTE/ASM, PCA instruments and MAXI/GSC data provided by RIKEN, JAXA and the MAXI team.
We also acknowledge Matsuoka et al. 2009 for all MAXI related work.
D.D. acknowledges support from the Govt. of West Bengal and DST/SERB sponsored Extra Mural Research project (EMR/2016/003918) fund.
S.N. acknowledges the support from a fellowship from Abdus Salam International Centre for Theoretical Physics, Italy, granted 
to Indian Centre for Space Physics, which allowed him to pursue the present research.

\end{document}